\documentclass[conference,a4paper]{IEEEtran}
\IEEEoverridecommandlockouts
\usepackage{cite}
\usepackage{amsmath,amssymb,amsfonts,bm}
\usepackage{algorithmic}
\usepackage{graphicx}
\usepackage{textcomp}
\usepackage{diagbox}
\usepackage{nicefrac}
\usepackage{comment}
\usepackage{xcolor}
\usepackage[T1]{fontenc}
\usepackage{tikz}
\usetikzlibrary{arrows.meta,calc}
\usepackage[nameinlink,capitalize]{cleveref}
\def\BibTeX{{\rm B\kern-.05em{\sc i\kern-.025em b}\kern-.08em
    T\kern-.1667em\lower.7ex\hbox{E}\kern-.125emX}}

\begin{document}\bstctlcite{IEEEexample:BSTcontrol}

\title{Robustness of Reinforcement Learning-Based Congestion Management in Low-Voltage Grids}

\author{
\IEEEauthorblockN{
Josef Hoppe$^{1}$\IEEEauthorrefmark{1}, 
Sarra Bouchkati$^1$\IEEEauthorrefmark{2}, 
Farah Nasr$^1$\IEEEauthorrefmark{2}, 
Jonathan Krapp\IEEEauthorrefmark{1},
Alexander Och\IEEEauthorrefmark{2},\\
Maximilian Wirth\IEEEauthorrefmark{3},
Jan Schiefelbein-Lach\IEEEauthorrefmark{4},
Oliver Pohl\IEEEauthorrefmark{5},  
Andreas Ulbig\IEEEauthorrefmark{2}, and
Michael T. Schaub\IEEEauthorrefmark{1}
}
  \IEEEauthorblockA{\IEEEauthorrefmark{1} Computational Network Science, RWTH Aachen University \\
  \{hoppe, schaub\}@netsci.rwth-aachen.de}
  \IEEEauthorblockA{\IEEEauthorrefmark{2} IAEW, RWTH Aachen University \\
   \{{s.bouchkati, f.nasr, a.och, a.ulbig}\}@iaew.rwth-aachen.de
   }
     \IEEEauthorblockA{\IEEEauthorrefmark{3} E.ON impulse GmbH}
  \IEEEauthorblockA{\IEEEauthorrefmark{4} E.ON Group Innovation GmbH}
  \IEEEauthorblockA{\IEEEauthorrefmark{5} Schleswig-Holstein Netz GmbH}

}

\maketitle

\def\thefootnote{1}\footnotetext{These authors contributed equally to this work}

\begin{abstract}
Increases in photovoltaic generation, charging of electric vehicles and heat-pump demand challenge operating limits in low-voltage distribution grids.
This requires curative curtailment methods that can operate under sparse observability, noisy measurements, and imperfect grid models. Unlike prior end-to-end reinforcement-learning approaches for partially observable curtailment, this work decouples congestion detection and control by combining a random-forest violation pre-classifier with an actor-critic controller, and evaluates its robustness to measurement noise and grid-parameter mismatch. The framework is tested on a real low-voltage grid using synthetic future operating scenarios with low observability and controllability.
With accurate grid parameters, the controller reduces total violation magnitude by 98.9\%, and this performance remains nearly unchanged under the tested measurement-noise settings. Grid-model mismatch proves to be more challenging, but the controller still mitigates most violations under the tested mismatch assumptions. 
\end{abstract}
\section{Introduction}

The decarbonization of energy systems is increasing the operational challenges faced by Distribution System Operators (DSOs). In low-voltage grids, the growing integration of photovoltaic (PV) systems, electric vehicles (EVs), and heat pumps is transforming grids originally designed for unidirectional power flow into systems with bidirectional and highly time-varying injections. Thus, operational limits such as thermal line loadings, voltage bounds, and transformer capacities are increasingly at risk of violation, making active congestion management essential for safe and reliable operation \cite{MEHMOOD2026127342}.

A common corrective action is the curtailment of controllable generation or demand to relieve asset overloading or voltage violations. The main technical challenge is to determine which asset should be adjusted, and by how much, so that grid constraints are satisfied while intervention costs are minimized. Due to volatile supply-demand conditions, such decisions must often be made in real time, within only a few minutes \cite{CAPITANESCU20111731}. Although this can be formulated through Optimal Power Flow (OPF) methods \cite{cain2012history,bienstock2019strong}, centralized OPF requires extensive measurement and communication infrastructure that is often unavailable or unreliable in practice.

Edge computing has therefore been investigated as a way to reduce communication overhead and avoid single points of failure by processing data closer to its source. However, practical deployment remains challenging because only a subset of grid buses is typically observable and controllable. Classical optimization-based approaches generally assume full state information, while state estimation relies on structural assumptions and pseudo-measurements that introduce additional uncertainty. Moreover, measurement errors and inaccurate grid parameters, such as uncertain line impedances caused by incomplete documentation and aging infrastructure, create a mismatch between the simulated model and the physical system \cite{ZHAN2025111468}. These limitations reduce the practical applicability of traditional optimization methods. In recent years, Machine Learning approaches have emerged as robust and computationally efficient alternatives to address these challenges \cite{LI2024122779}.

\subsection{Related Work}
Several studies have been proposed for approximating the AC-OPF problem for congestion management. In particular, Reinforcement Learning (RL) has been explored as a promising alternative for smart grid control \cite{marchesini2025rl2gridbenchmarkingreinforcementlearning}.

For this, it is important to integrate constraints into the control model.
Some approaches work by modifying reward functions to integrate further constraints directly \cite{achiam2017constrained,CAO2021rewardcnstraint}, especially by introducing system-specific Lagrangian penalties of constraint violations \cite{yan2020,TongRL}.
To provide more rigorous physical feasibility guarantees, \cite{sayedRLFeasibility} instead introduces a safe convex layer, while \cite{SayedRL2} utilizes a holomorphic embedding layer within the neural architecture. Further advancing physics-informed RL, \cite{Wu2025PIRL} proposed a framework based on power flow equations that augments the policy gradient with constraints to correct infeasible generation operations. 
Other approaches focus on post-processing the agent's actions to ensure validity: \cite{Zhou2020DRL} enforces generation limits using a power flow solver, while \cite{Yi2023KDRL} employs linear programming and action adjustment techniques to correct infeasible setpoints.

Despite these advances, a common limitation is the assumption of a fully observable grid state. 
In practical settings, full observability is rarely achieved due to slow smart meter rollout or restricted access to consumer data. 
While our previous work \cite{Wolf2024Reinforcement} presented an end-to-end RL approach tailored for partial observability, the impact of real-world uncertainties remains a critical concern. 
Inaccuracies in grid models and measurement errors can jeopardize the performance and reliability of RL agents. 
Furthermore, scalability remains a significant challenge, as end-to-end frameworks must simultaneously navigate two complex tasks: identifying constraint violations under limited observability and determining optimal setpoints. 
The introduction of constraints already shows different approaches for handling this complexity, including separating the tasks into multiple steps, which can also be applied to the divide between detecting violations and solving them precisely.

\subsection{Contributions}
In this work, we propose a framework for the uncertainty-aware evaluation of RL-based voltage and congestion control in low-voltage distribution grids. 
Building upon the actor–critic architecture introduced in \cite{Wolf2024Reinforcement}, we decouple the control logic by integrating a random forest classifier for congestion prediction, reducing the complexity of the learning task and enhancing scalability. Furthermore, we assess the robustness of the framework against model mismatch and measurement noise.
The main contributions of this work are:

\begin{itemize}
    \item a modular two-step control pipeline that combines a random-forest violation pre-classifier with an actor-critic controller under partial observability,
    \item an empirical evaluation framework for assessing sensitivity to parametric model mismatch and measurement noise,
    \item a metric set to jointly evaluate violation mitigation and curtailment side effects, and
    \item a case study on a real low-voltage grid topology with synthetic operating situations.
\end{itemize}

\section{Problem Formulation}

This section defines the power curtailment problem. The goal is to determine the optimal operating setpoints of controllable units, while accounting for the grid constraints and the requirements of edge-based operation: partial observability and system uncertainties.

\subsection{Grid Model and AC-OPF}
We consider a low-voltage distribution grid represented as a graph $\mathcal{G} = (\mathcal{V}, \mathcal{E})$, where $\mathcal{V}$ and $\mathcal{E}$ denote the sets of buses and branches, respectively. For each bus $i \in \mathcal{V}$, the complex voltage is $V_i = V_{m,i}e^{j\theta_i}$, where $V_{m,i}$ denotes the voltage magnitude and $\theta_i$ denotes the voltage angle at bus $i$. The apparent power $S_i$ is determined partially by controllable DER generation $S_{g,i} = P_{g,i} + jQ_{g,i}$ and partially by flexible load demand $S_{d,i} = P_{d,i} + jQ_{d,i}$.
Bold symbols denote a vector over the entire grid that consists of the individual (non-bold) measurements attached to buses or branches, respectively.

The objective of active congestion management is to minimize the cost of intervention (curtailment) while satisfying the nonlinear AC power flow equations and operational limits.
The generation (consumption) at bus $i\in\mathcal{V}$ is curtailed by $P_{g,i}^{C}$ ($P_{d,i}^{C}$).
In this paper, we set the curtailment costs $\alpha_i$ for generation and $\beta_i$ for consumption $\alpha_i=\beta_i=1$ for all buses, assuming the total curtailment as cost.
This is formulated as an AC-OPF problem with:
\begin{subequations}
\label{eq:unified_opf}
\begin{align}
    \min_{\mathbf{u}} \quad & f(\mathbf{u}) = \sum_{i \in \mathcal{V}} \alpha_i P_{g,i}^C + \beta_i P_{d,i}^C \label{eq:opf_obj} \\
    \text{s.t.} \quad & \mathbf{S}_{\mathrm{bus}}(\mathbf{V},  \bm{\gamma}) = \mathbf{P}(\mathbf{u}) + j\mathbf{Q}(\mathbf{u}), \label{eq:pf_equil} \\
    & V_m^{\min} \leq V_{m,i} \leq V_m^{\max}, \quad \forall i \in \mathcal{V} \label{eq:volt_const} \\
    & |S_{ij}(\mathbf{V},  \bm{\gamma})| \leq S_{ij}^{\max}, \quad \forall (i,j) \in \mathcal{E} \label{eq:therm_const} \\
    & 0 \leq P_{g,i}^C \leq P_{g,i}^{C,\mathrm{max}}, \quad 0 \leq P_{d,i}^C \leq P_{d,i}^{C,\mathrm{max}} \label{eq:ctrl_limits}
\end{align}
\end{subequations}

where $\mathbf{u} = [\bm{P}_g^C, \bm{P}_d^C]^\top$ is the vector of control actions, and $\bm{\gamma}$ represents the grid parameters (e.g., line impedances). Equation \eqref{eq:pf_equil} represents the physical power balance, while \eqref{eq:volt_const} and \eqref{eq:therm_const} ensure minimum ($V_m^{\min}$) and maximum ($V_m^{\max}$) voltage magnitude  and thermal constraints (as maximum apparent branch power $S_{ij}^{\max}$).
\eqref{eq:ctrl_limits} defines the maximum curtailment per bus; to define a bus as not controllable, we set maximum curtailment to zero ($P_{g,i}^{C,\max}=P_{d,i}^{C,max}=0$).

\subsection{Partial Observability and Uncertainty}
In practical edge deployments, the controller lacks access to the full system state $\mathbf{x} = [\mathbf{P}, \mathbf{Q},\mathbf{V}_m, \bm{\theta}]$. Instead, the agent operates under partial observability. Let $\mathcal{V}_{\mathrm{obs}} \subset \mathcal{V}$ be the set of observable buses. The observation vector $\mathbf{z}$ is defined as:
\begin{equation}
\mathbf{z} = \mathbf{h}(\mathbf{x}_{\mathrm{obs}}) + \bm{\eta}
\end{equation}
where $\mathbf{h}(\cdot)$ is the observation function for the monitored subset and $\bm{\eta}$ represents measurement noise (e.g., sensor inaccuracies or synchronization errors). Furthermore, the physical laws in \eqref{eq:pf_equil} and \eqref{eq:therm_const} depend on the vector $\bm{\gamma}$ of grid parameters (e.g. impedances), which  is subject to model uncertainty.
We represent this uncertainty as a relative parametric mismatch,
\begin{equation}
    \bm{\gamma}_{\mathrm{mis}}
    =
    \bm{\gamma}_{\mathrm{nom}} \odot ( \bm{1}+\Delta\bm{\gamma}),
\end{equation}
where $\bm{\gamma}_{\mathrm{nom}}$ denotes the nominal grid model and $\Delta\bm{\gamma}$ represents the relative parameter mismatch.
The challenge addressed in this work is to derive a control policy $\pi: \mathbf{z} \to \mathbf{u}$ that minimizes \eqref{eq:opf_obj} and satisfies constraints \eqref{eq:volt_const}--\eqref{eq:therm_const} with noisy, incomplete observations $\mathbf{z}$ and an inaccurate model $\bm{\gamma}$.

\section{Methodology}
This section outlines the methodology used to develop and train a model for grid curtailment with partial measurement availability. The model is capable of handling various operational scenarios ranging from non-critical to critical grid states.

\subsection{Dataset Generation}\label{sec:dataset}
\begin{figure}[t]
    \centering
    \includegraphics[width=0.85\linewidth]{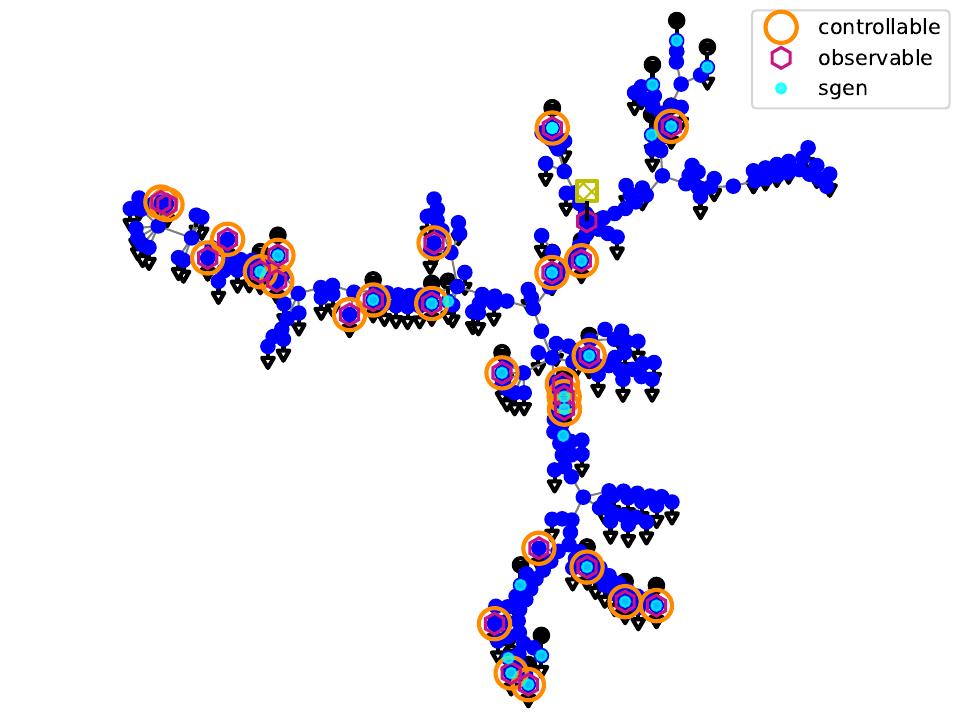}
    \caption{Low Voltage grid provided by Schleswig-Holstein Netz GmbH; with randomly assigned controllable and observable buses.}
    \label{fig:net}
\end{figure}
The dataset is derived from the topology of a real-world low-voltage distribution grid provided by Schleswig-Holstein Netz AG. It represents a future operating scenario with increased penetration of PV systems, EVs, and heat pumps. Flexible units serve as controllable grid connection points for congestion management: their active power can be adjusted through external control signals within predefined technical limits. PV systems provide flexibility through downward curtailment of power generation, whereas EV charging and heat pumps provide demand-side flexibility by adjusting their power consumption.

Following the regulatory framework in Germany, 10\% of the available buses are defined as both observable and controllable (see \cref{fig:net}). Time-series profiles are assigned to all observable and unobservable connection points using future scenarios from SimBench \cite{meinecke2020simbench}. A quasi-static time-series simulation with 15-minute resolution over one year yields 35\,040 operating points. For each operating point, the full grid state is calculated using an AC power-flow (PF) calculation. Results exceeding predefined physical limits, specifically voltage magnitudes outside $V_m \notin [0.8, 1.2]$ p.u., are excluded, as such extreme conditions would typically be addressed by protection systems rather than operational congestion management.

The PF results are used as the baseline, while OPF solutions provide a reference for comparison. During training, however, the model receives only PF-based information and no OPF solution data. In contrast to \cite{Wolf2024Reinforcement}, dataset augmentation is not required because the improved control model is trained almost exclusively on operating points with violations.

\subsection{Uncertainty Modeling}

In this work, we consider two types of uncertainties:

\paragraph{Grid Model Mismatch}

As discussed above, grid parameter uncertainty may arise from inaccurate documentation, aging infrastructure, and imperfect knowledge of line parameters. In contrast to works that use uniformly distributed perturbations to assess bounded worst-case mismatch~\cite{Jouini2021}, we model inaccuracies as relative Gaussian perturbations, which are usually small but may occasionally be larger.

For each line and each parameter e.g. resistance, reactance, and capacitance per unit length, we define
\begin{equation}
    \gamma_{\mathrm{mis}}
    =
    \max\left(\gamma_{\mathrm{nom}}(1+\epsilon_{\gamma}),\,0\right),
    \qquad
    \epsilon_{\gamma} \sim \mathcal{N}(0,\eta_{\mathrm{mis}}^2),
\end{equation}
where $\eta_{\mathrm{mis}}=0.05$. Perturbations are sampled independently for each line and parameter, and clipping ensures non-negative physical values. The resulting mismatched grid represents the imperfect model available to the controller, while the original grid serves as the reference physical system.

\paragraph{Measurement Uncertainty}
Similar to previous work \cite{wen2026heterogeneous,dabush2023state}, we model measurement noise by adding zero-mean Gaussian noise to the measured quantities before feeding them to the observations.
The resulting action is applied to the true (noise-free) system state to isolate the effect of observation errors on decision-making.
The standard deviation of the noise is equal to half of the uncertainty range as quantified by the manufacturer $\bm{\eta}\sim \mathcal{N}(0,\sigma^2) $.
To analyze the robustness more clearly, we use different multipliers for the noise, increasing or decreasing the standard deviation proportionally.

\subsection{Control Model}
\label{sec:control_rl}

\begin{figure}[t]
    \centering
    {\large Evaluation}
    \includegraphics[width=\linewidth]{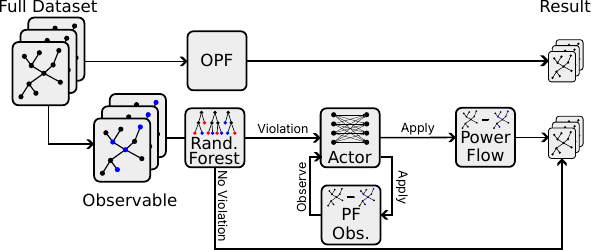}
    \caption{Flow of the Evaluation. For cases where the pre-classifier predicts a violation, the actor generates multiple control steps (with updated observations) to solve the violation. Otherwise, we collect the original grid state as the result. We calculate the Optimal Power Flow (OPF) from the full data.}
    \label{fig:eval-workflow}
\end{figure}

For each operating point, the control model is executed for a fixed number of control steps. The control model consists of a random forest classifier that classifies observations into violations or non-violations, and a reinforcement-learning actor that learns to minimally curtail power to resolve violations.

\subsubsection{Random Forest pre-classifier}
In order to detect if a given observation contains a violation, we fit a random forest classifier to classify load situations into violations or non-violations.
This reduces the complexity of the target learned by the RL agent and speeds up the evaluation by reducing the number of time steps that need to be simulated.

\subsubsection{RL Agent}
The control policy is parameterized by a multilayer perceptron (MLP), called the \emph{actor}, that maps observable measurements to curtailment actions. The input consists of measurements from observable buses; the output is a vector of curtailment factors for the controllable buses.

\subsection{Training Setup}

For training and evaluation, we split the cases in the dataset into 60\% training data, 20\% validation data, and 20\% test data.
The RL agent and the random forest are trained on the training data; and we select an agent based on the performance on the validation data.
We report the results obtained on the test set to avoid erroneous good performance from the selection of the agent.
The split is stratified by the type of violation to ensure each of the three sets is representative of the overall dataset.
For this, the dataset is first split into the three occurring types of violation (no violation, lower voltage band violation, combination voltage band and branch load violation).
Then, each set is split separately according to the above proportions, and the resulting subsets are combined to form the final training, validation, and test sets.

\begin{figure}[t]
    \centering
    {\large Training Setup}\\\vspace{3pt}
    \includegraphics[width=\linewidth]{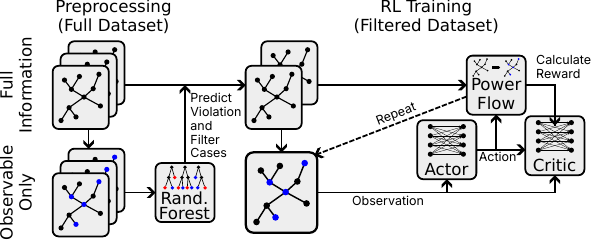}
    \caption{Training setup. A Random Forest filters non-violations; the RL loop takes cases, evaluates actions and rewards, and updates the actor-critic weights.
}
    \label{fig:train-env}
\end{figure}

We first train the random forest classifier on the training set.
Then, we train the RL agent on all cases classified as violations by the random forest classifier (see \cref{fig:train-env}).
The policy is trained using an actor–critic reinforcement learning algorithm with deep deterministic policy gradient \cite{tan2021reinforcement,sumiea2024deep}.
Training consists of two parts: Collecting observations and training the actor and critic networks.
The reinforcement learning algorithm obtains observations by first applying the actor to the observable measurements, obtaining an action.
To improve variety in early observations, actions are modified with decreasing amounts of noise before being applied.
The action is then applied and the algorithm calculates the power flow and calculates the reward from the full grid state.
One observation is the triple of input, action, and reward.
The critic is an MLP that is trained to learn the reward from the input and the action.
For the training of the actor, the observations and output of the actor are connected to the critic network.
Then, without updating the critic weights, the actor is trained with gradient descent, maximizing the output of the critic.
We refer the reader to \cite{tan2021reinforcement} for further details.

The reward function penalizes violations of operating parameters, or, if none exist, rewards smaller curtailment:
\begin{align}
    r &= \begin{cases}
        -\nu (\sqrt{V_{\text{err}} / 0.2} + \sqrt{S_{\text{err}}}) &\text{if violation}\\
        1 - \frac{\sqrt{\sum_{i\in\mathcal{V}} P_{g,i}^C + P_{d,i}^C}}{\lambda \sqrt{\sum_{i\in\mathcal{V}} P_{g,i}^{C,\max} + P_{d,i}^{C,\max}}}&\text{else}%
        \end{cases}\\
    V_{\text{err}} &= \max_{i\in\mathcal{V}} \max \{V_{m}^{\max} - V_{m,i}, V_{m,i} - V_{m}^{\min}, 0\}\\
        S_{\text{err}} &= \max_{(i,j)\in\mathcal{E}}\max\left\{ \frac{|S_{i,j}|}{S_{i,j}^{\max}} - 1, 0\right\}
\end{align}
The voltage error $V_{\text{err}}$ (maximum absolute violation) and branch load error $S_{\text{err}}$ (maximum relative overload) are scaled to realistic maximum violations of $0.2$ p.u. and +100 \%, respectively.
The hyperparameters $\nu$ and $\lambda$ tune the relation between positive and negative rewards and the gap between small violations and maximum curtailment.
It should be noted that the full grid state is only used during offline training and evaluation for reward calculation. In this stage, the simulator provides access to all bus voltages and branch loadings, which allows violation labels and rewards to be computed from the complete system state. This information is not required during deployment: once trained, the random-forest classifier and actor receive only the partial, noisy observation vector from the observable buses. 

To allow iterative correction, the resulting state can be re-evaluated by the actor for multiple control steps.
Both during and after training, we apply the model as follows:
\begin{enumerate}
    \item A new observation is taken from observable buses.
    \item The random forest classifier classifies the observation as \emph{violation} or \emph{no violation}.\vspace{-2pt}
    \begin{itemize}
        \item If the observation is classified as \emph{no violation}, take no action. Wait for next observation and go to 1).\vspace{-2pt}
    \end{itemize}
    \item The actor generates an action from the observation.
    \item Apply the action to the controllable elements of the grid. Wait for the next observation and go to 3).
\end{enumerate}

\subsection{Evaluation Metrics}

We quantify the overall performance using a set of metrics. Specifically, we quantify the reduction in violation magnitude (compared to the baseline with no action), the number of remaining cases with violations, and the amount of curtailment (compared to the OPF) needed to achieve it.
For the pre-classifier, we also report the number of false negatives and false positives in detecting violations.

\section{Experiments}

All results reported here stem from experiments on the dataset based on the low-voltage grid provided by Schleswig-Holstein Netz GmbH as described in \cref{sec:dataset}.
The hyperparameters of the selected runs are $\lambda=1.07$, $\nu=0.98$, with 5 layers of width 64 and 200k control steps for the training without pre-classification; $\lambda=1.14$, $\nu=0.77$, with 8 layers of width 32 and 500k control steps for the normal dataset; and $\lambda=1.18$, $\nu=0.33$, with 5 layers of width 128 and 500k control steps for the mismatch dataset.

\begin{figure}
    \centering
    \includegraphics[width=\linewidth]{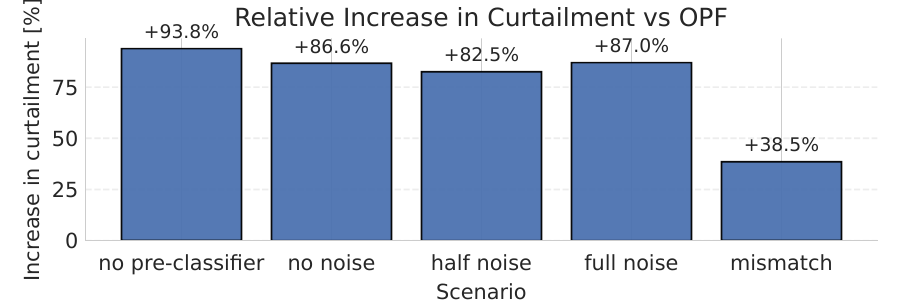}
    \caption{Excess curtailment of actors, relative to OPF curtailment. Different scenarios, each shows the total over cases with and without violations.}
    \label{fig:curtailment}
\end{figure}

\begin{figure}
    \centering
    \includegraphics[width=\linewidth]{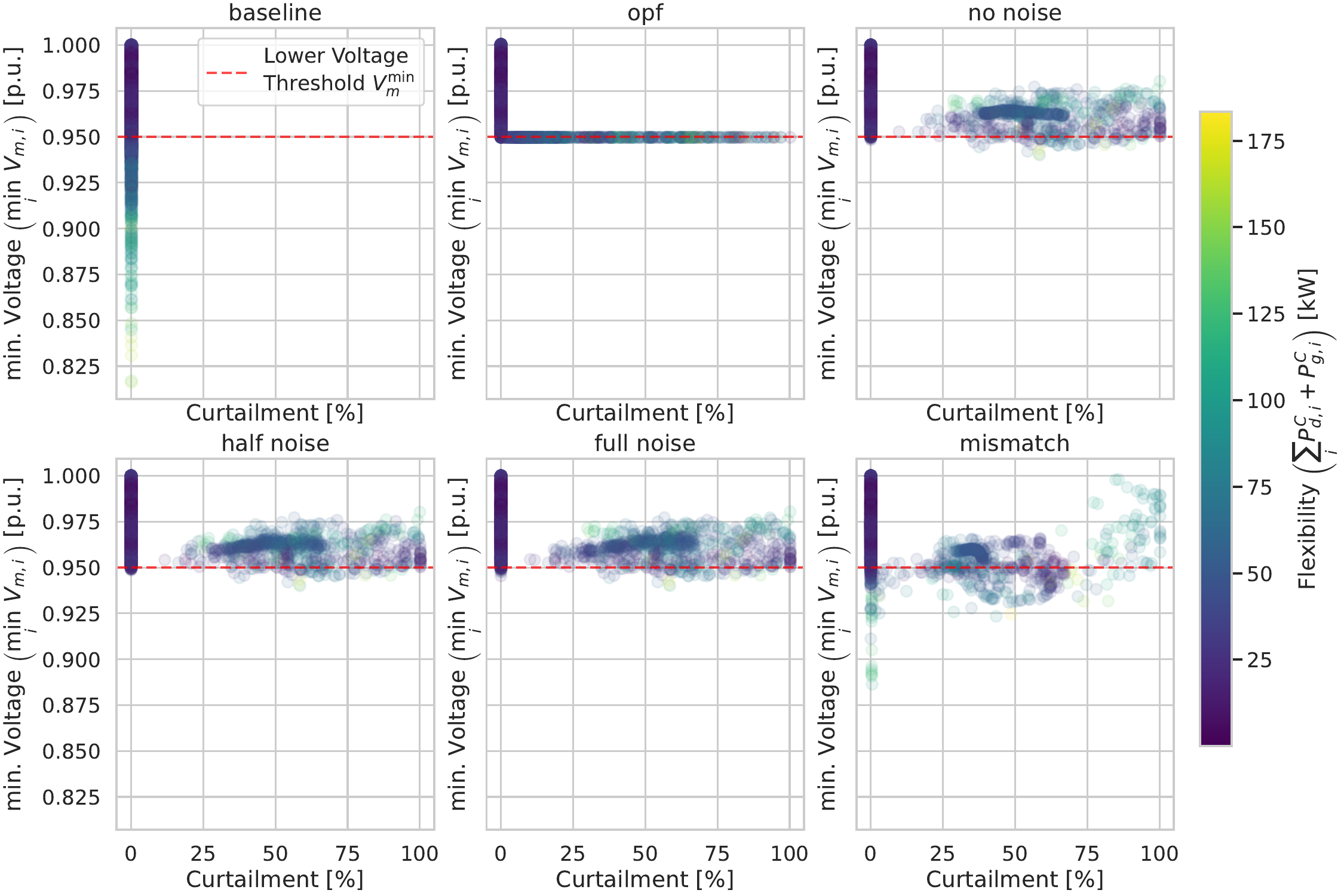}
    \caption{Lowest Voltage and curtailment, different solutions and scenarios. On the x axis, the plot shows the amount of curtailment (relative to maximum flexibility). The y axis shows the smallest voltage in the overall grid. Colored by total amount of flexibility for visual comparison.}
    \label{fig:scatter}
\end{figure}

\subsection{Implementation}

We use the scikit-learn implementation of a random forest classifier for the pre-classification.
The actor-critic model and training is implemented using PyTorch and TorchRL \cite{boutorchrl}.
The training data is stored in pypower \cite{lincoln_pypower} case format.

We train on the training subsets of both the original dataset and the mismatch dataset without measurement noise.
We perform a hyperparameter search with random exploration for a total of 20 runs and evaluate the agents on the validation subset of the dataset they were trained on.
Then, we select the best run based on the magnitude of violations after applying the action and the total amount of power curtailed.

For the evaluation, we run both original and mismatch agents on the evaluation subset of the original dataset.
In the presence of measurement noise, we modify the observations, applying the action to the noise-free ground truth.
This results in directly comparable metrics across evaluations.

\subsection{Results}

\begin{table}[t]
    \caption{Number of false positive and negative predictions of violations by the random forest classifier.} 
    \centering
    \begin{tabular}{l|rrrr}
        Noise Factor & $0.0$ & $0.5$ & $1.0$ & Mismatch\\
        \hline
        False Pos & 7 & 20 & 49 & 4 \\
        False Pos (\%) & 0.2 \% & 0.6 \% & 1.4 \% & 0.1 \% \\
        False Neg & 4 & 10 & 15 & 119 \\
        False Neg (\%) & 0.3 \% & 0.7 \% & 1.2 \% & 9.2 \% \\
    \end{tabular}

    \label{tab:rf-classification}
\end{table}

\begin{table}[t]
    \caption{Number and magnitude of violation in different scenarios.}
    \centering
    \begin{tabular}{l|lll}
    & Mean  & Violation & Number of \\
    & Violation   & Magnitude & Violations \\
    & [p.u.]        & Reduction & Remaining \\
    \hline
    Base       & 3.0E-3  & \rule{0mm}{2mm} 0\% & \rule{10mm}{2mm} 726 \\
    No Noise   & 3.2E-5 & \rule{9.89mm}{2mm} 98.9\% & \rule{0.83mm}{2mm} 60 \\
    Half Noise & 3.3E-5 & \rule{9.89mm}{2mm} 98.9\% &  \rule{0.9mm}{2mm} 65 \\
    Noise      & 3.5E-5 & \rule{9.88mm}{2mm} 98.8\% & \rule{0.96mm}{2mm} 70 \\
    Mismatch   & 4.6E-4 & \rule{7.96mm}{2mm} 79.6\% &  \rule{4.1mm}{2mm} 301 \\
    No Pre-Classifier  & 4.6E-4 & \rule{9.09mm}{2mm} 90.9\% &  \rule{2.4mm}{2mm} 176 \\
\end{tabular}
    \label{tab:combined}
\end{table}

We first look at the effect of the pre-classifier, then different levels of noise and finally consider  grid model mismatch.

\subsubsection{Two-step Control Model}
With the same number of training steps, the slightly-modified end-to-end approach \cite{Wolf2024Reinforcement} curtails more total power than the two-step approach (cf. \cref{fig:curtailment}) while reducing the total violation by less and resolving fewer cases with violations completely (cf. \cref{tab:combined}).
There are both unnecessary control actions without violations and insufficient curtailment in operating points with large violations.

\subsubsection{Results with Accurate Measurements and Model}

First, we look at the best-case results obtained with partial observability.
\Cref{fig:curtailment} shows that with accurate partial information, the actor curtails less than twice the amount that the OPF does.
With this, the actor is able to reduce the amount of violation by almost $99\%$, solving most cases (see \cref{tab:combined}).
\Cref{fig:scatter} gives a more visual overview of this result, showing a large group of cases that are solved with a small margin to the defined minimum voltage, and a bigger group with varying margins.

Other runs with other random seeds that we considered provided alternative trade-offs, either increasing excess curtailment and decreasing remaining violations, or vice versa.

\subsubsection{Measurement Noise}

Second, we look at the accuracy of the approach in the presence of noise.
\Cref{tab:rf-classification} shows the performance of the random forest classifier on the test set for different noise levels.
While its accuracy drops in the presence of faulty measurements, the overall number of misclassifications is very small.
\Cref{tab:combined} shows that the combined model reduces the violations by 98.9 \%, and is robust to noise.

\subsubsection{Grid Model Mismatch}

In the presence of model mismatch, the pre-classifier has substantially more false negatives (see \cref{tab:rf-classification}) and \cref{fig:curtailment} shows a smaller total curtailment for the control model.
In \cref{tab:combined}, we see a spike in the number of cases with violations after the control action is applied, but the overall magnitude of violations is still reduced by almost 80 \%.
This indicates that the mismatch is skewed towards producing higher minimum voltages in the grid, biasing the control model toward smaller curtailment.
\Cref{fig:scatter} confirms this, with the bulk of control actions having slightly less curtailment.
We also see two more extreme clusters, one with a bigger overcurtailment and one with a number of cases that have no curtailment and relatively strong constraint violations.
Overall, the vast majority of misclassifications is close to the threshold, as is the vast majority of remaining violations.
\section{Conclusion}
This work presents an improved two-step control model and evaluates its robustness to measurement noise and grid model mismatch, with the overall goal of improving the real-world viability of ML-based congestion management.

In the experiments, the two-step model clearly outperforms an end-to-end model for the same number of training steps. Although additional training or a more extensive hyperparameter search could likely improve the end-to-end baseline, the results show a clear computational advantage for the two-step approach. Since such models must be deployed at scale and periodically retrained, reduced training cost is an important factor for practical applicability.

The results further show that reinforcement-learning-based curative curtailment can remain effective in low-voltage distribution grids even when only a small share of buses is observable and controllable. By separating violation detection through a random-forest pre-classifier from curtailment control through an actor-critic controller, the proposed framework reduces the complexity of the learning task while maintaining strong control performance. On a real low-voltage grid, the controller reduces total violation magnitude by 98.9\% under accurate grid parameters and achieves almost identical performance under measurement noise. This indicates that realistic measurement uncertainty is not the main limiting factor for edge-based congestion management.
By contrast, grid-model mismatch has a stronger impact. Under mismatched grid parameters, the violation-magnitude reduction decreases to 79.6\%, and the number of remaining violation cases increases substantially. Nevertheless, the controller still reduces violation severity compared to the uncontrolled baseline. Thus, imperfect model knowledge does not invalidate learning-based curtailment, but shifts the practical objective from exact optimality toward robust mitigation. For DSOs, this distinction is important: a controller that substantially reduces unseen violations may still be valuable in operation if model validation and safety safeguards are included.

Overall, the case study shows that some likely real-world errors do not lead to catastrophic performance degradation, while the reduced computational cost improves the practical applicability of the proposed control framework. Future work should extend the robustness analysis to additional sources of mismatch, including topology uncertainty, changing load and generation profiles, and errors in the assumed availability of flexible units. Further important directions include online model calibration or adaptive retraining, as well as evaluation on multiple grid topologies and, where possible, with field data or hardware-in-the-loop experiments to assess reliability under operational constraints, communication delays, and regulatory requirements for curtailment.

\bstctlcite{IEEEexample:BSTcontrol}

\bibliographystyle{IEEEtran}
\bibliography{IEEEabrv,Bibliography}

\end{document}